\documentclass{cimento}
\usepackage{graphicx}
\usepackage{dcolumn}
\usepackage{multirow}
\usepackage{amssymb,amsmath}

\newcommand{\SNR}{\text{SNR}}

\newcommand{\HW}{H_{\text{NSD}}}

\newcommand{\bra}[1]{\langle #1|}
\newcommand{\ket}[1]{| #1\rangle}
\newcommand{\CG}[3]{\langle #1;#2|#3\rangle}

\newcommand{\evec}{\boldsymbol{\epsilon}}

\newcommand{\kvec}{\mathbf{k}}

\newcommand{\Efield}{\mathbf{E}}
\newcommand{\Bfield}{\mathbf{B}}

\title{Atomic parity violation in two-photon $J=0\rightarrow 1$ transitions}
\author{
D.~R.~Dounas-Frazer\from{ins:UCB}\ETC\thanks{drdf@berkeley.edu},
K.~Tsigutkin\from{ins:UCB}, D.~English\from{ins:UCB} \atque
D.~Budker\from{ins:UCB}\from{ins:LBNL}} \instlist{ \inst{ins:UCB}
Department of Physics, University of California at Berkeley,
Berkeley, California 94720-7300, USA \inst{ins:LBNL} Nuclear Science
Division, Lawrence Berkeley National Laboratory, Berkeley,
California 94720, USA}

\PACSes{
\PACSit{11.30.Er}{Charge conjugation, parity, time reversal, and other discrete symmetries}
\PACSit{32.80.Rm}{Multiphoton ionization and excitation to highly excited states}
}
\begin{document}

\maketitle

\begin{abstract}
We present a method for measuring nuclear-spin-dependent atomic parity violation without nuclear-spin-independent background. Such measurements can be achieved by observing interference of parity-conserving and parity-violating two-photon $J=0\rightarrow 1$ transitions driven by collinear photons of the same frequency in the presence of an external static magnetic field. \end{abstract}

Sub-1\% measurements of nuclear-spin-independent (NSI) atomic parity
violation (APV) \cite{Wood1997, Bennett1999} have led to precise evaluation of the nuclear weak charge \cite{Porsev2009}, yielding excellent agreement with the Standard Model at low energies. On the other hand, nuclear anapole moments, parity-violating moments induced by weak interactions within the nucleus, have been extracted from nuclear-spin-dependent (NSD) APV measurements in Cs~\cite{Flambaum1997} and Tl~\cite{Kozlov2002}. In these experiments, NSD APV was observed as a small correction to NSI effects~\cite{Flambaum1980}. The values of the Cs and Tl anapole moments result in constraints on weak coupling constants that are difficult to reconcile with those obtained from other nuclear physics experiments and with each other~\cite{Haxton2002,Ginges2004}. Future anapole moment measurements will provide additional insight into this open problem, and are a major goal of ongoing experiments in Yb~\cite{Tsigutkin2009}, Dy~\cite{Nguyen1997}, Fr~\cite{Gomez2007}, Ra$^{+}$~\cite{DzubaPRA2011,Giri2011}, and diatomic molecules~\cite{DeMille2008}. In this work, we propose a technique that allows for measurement of NSD APV without NSI background in atoms with nonzero nuclear spin.

The proposed method uses two-photon transitions from an initial state of total electronic angular momentum $J_i=0$ to an opposite-parity $J_f=1$ final state (or \emph{vice versa}). The APV signal is due to interference of parity-conserving electric-dipole-electric-quadrupole ($E1$-$E2$) and electric-dipole-magnetic-dipole ($E1$-$M1$) transitions with parity-violating $E1$-$E1$ transitions induced by the weak interaction. This scheme is different from other multi-photon APV schemes~\cite{DounasFrazer2011,GunawardenaPRL2007,Guena2003,Cronin1998} in that
the transitions are driven by collinear photons of the same frequency, and hence are subject to a Bose-Einstein statistics (BES) selection rule that forbids $E1$-$E1$ $J=0\rightarrow 1$ transitions~\cite{DeMille2000, English2010}. However, such transitions may be induced by perturbations that cause the final state to mix with opposite-parity $J\neq 1$ states, such as the NSD weak interaction and, in the presence of an external static electric field, the Stark effect. Because the NSI weak interaction only leads to mixing of the final state with other $J=1$ states, it cannot induce $J=0\rightarrow 1$ transitions. Thus NSI-background-free measurements of NSD APV can be achieved by exploiting two-photon BES selection rules. We call this method the \emph{degenerate photon scheme (DPS)}.

Consider atoms illuminated by linearly polarized light in the presence of a static magnetic field $\Bfield$. The optical field is characterized by polarization $\evec$, propagation  vector $\kvec$, frequency $\omega$, and intensity $\mathcal{I}$. We choose the frequency to be half the energy interval  $\omega_{fi}$ between the ground state $\ket{i}$ and an excited state $\ket{f}$ of opposite  nominal parity. We work in atomic units: $\hbar=|e|=m_e=1$. The transition rate satisfies $R= (4\pi\alpha)^2\mathcal{I}^2|A|^2/\Gamma$, where $\alpha$ is the fine structure constant, and $A$ and $\Gamma$ are the amplitude and width of the transition~\cite{Faisal1987}. Energy eigenstates are represented as $\ket{i}=\ket{J_i I  F_i M_i}$, and likewise for $\ket{f}$. Here $J_i$, $I$, and $F_i$ are quantum numbers associated with the electronic, nuclear, and total angular momentum, respectively, and $M_i\in\{\pm F_i,\pm(F_i-1),\hdots\}$ is the projection of $F_i$ along the quantization axis ($z$-axis), which we choose along $\Bfield$.

\begin{figure}
\includegraphics[width=\textwidth]{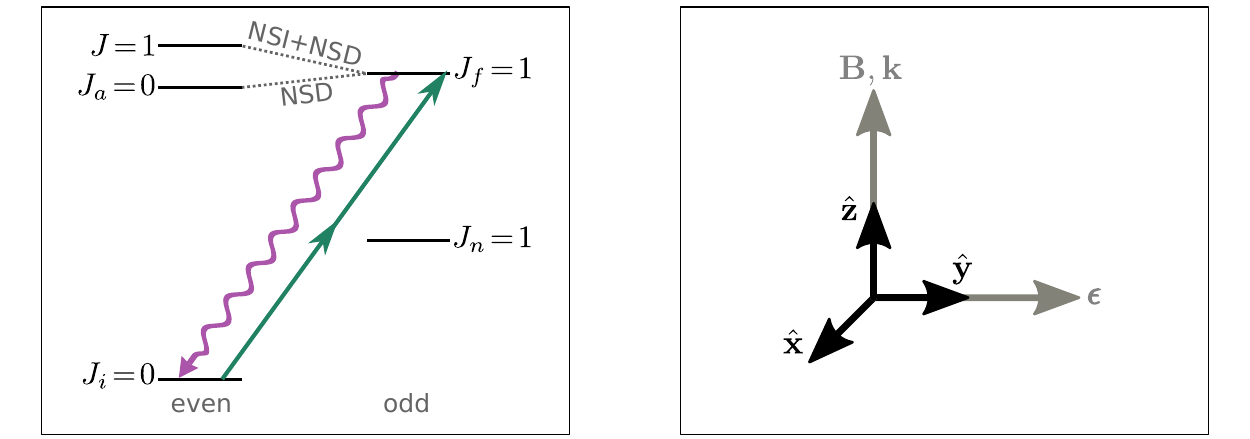}
\caption{\label{fig:1} (Left) Energy level diagram.  Dotted lines indicate APV mixing of opposite parity states, and upward- and downward-pointing arrows represent two-photon absorption and one-photon fluorescence, respectively. (Right) Field geometry. The propagation vector $\mathbf{k}$ may alternatively be anti-aligned with the magnetic field $\Bfield$.}
\end{figure}

The transition is enhanced by the presence of an intermediate state $\ket{n}$ of total electronic angular momentum $J_n=1$ whose energy lies about halfway between the energies of the initial and final states~(fig.~\ref{fig:1}). For typical situations, the energy defect $\Delta = \omega_{ni}-\omega_{fi}/2$ is large compared to the Rabi frequency $\Omega_\text{R}$ associated with the one-photon resonance involving the intermediate state. We assume that the scattering rate from $\ket{n}$ to $\ket{i}$ is small compared to the natural width $\Gamma_f$ of $\ket{f}$: $(\Omega_\text{R}/\Delta)^2\Gamma_n \ll \Gamma_f$. In this case, the system reduces to a two-level system consisting of initial and final states coupled by an effective optical field.

The parity-violating $E1$-$E1$ transition is induced by mixing of the final state with opposite-parity states via the weak interaction. In general, $\ket{f}$ may mix with states of electronic angular momentum $J = 0,  1$, or 2 according to the selection rules for NSD APV mixing~\cite{Khriplovich1991}. Mixing of the final state with  $J=1$ states results in a perturbed final state with electronic angular momentum~1 that cannot be excited via $E1$-$E1$ transitions. We limit our discussion to mixing of $\ket{f}$ with $J=0$ states (the case of mixing with $J=2$ states is similar), and  assume that this mixing is dominated by a single state $\ket{a}$ of total electronic angular momentum $J_a=0$. In this case, only transitions for which $F_f=I$ may be induced by the weak interaction. Transitions to hyperfine levels $F_f=I\pm1$ that arise due to parity-conserving processes can be used as  APV-free references, important for discriminating APV from systematic effects.

The amplitude for a degenerate two-photon $J=0\rightarrow 1$ transition is~\cite{Dounas-Frazer2011arXiv}:
\begin{equation}\label{eq:A}
A = i\mathcal{Q}k_{-q}(-1)^q\CG{F_iM_i}{1q}{F_fM_f}+
i\zeta\delta_{F_fF_i}\delta_{M_fM_i},
\end{equation}
where the terms proportional to $\mathcal{Q}$ and $\zeta$ are the amplitudes of the parity- conserving and violating transitions, respectively. Here $q=M_f-M_i$ is a spherical index, $k_q$ is the $q$th spherical component of $\hat{\kvec}$, $\langle F_iM_i;1q |F_fM_f\rangle$ is a Clebsch-Gordan coefficient, and $\delta_{F_fF_i}$ is the Kronecker delta. The quantities $\mathcal{Q}$ and $\zeta$ are
\begin{align}\label{eq:xi}
\mathcal{Q} = \frac{Q_{fn}d_{ni}}{2\sqrt{15}\Delta} + \frac{\mu_{fn}d_{ni}}{3\sqrt{2}\Delta}
\quad\text{and}\quad
\zeta =
\frac{\Omega_{fa}\,d_{an}\,d_{ni}}{3\omega_{fa}\Delta},
\end{align}
where the reduced matrix elements $Q_{fn}=(J_f||Q||J_n)$, $\mu_{fn}=(J_f||\mu||J_n)$, and $d_{ni}=(J_n||d||J_i)$ of the  electric quadrupole, magnetic dipole, and electric dipole moments, respectively, are independent of $F_f$ and $I$. Here  $\omega_{fa}=\omega_f-\omega_a$ is the energy difference of states $\ket{f}$ and $\ket{a}$, and $\Omega_{fa}$ is related to the matrix element of the NSD APV Hamiltonian $\HW$ by $\bra{f}\HW\ket{a}=i\Omega_{fa}$. The parameter $\Omega_{fa}$ must be a purely real quantity to preserve time reversal invariance~\cite{Khriplovich1991}.

The goal of the DPS is to observe interference of parity- violating and conserving amplitudes in the rate  $R$. When $M_f=M_i$, $R$ consists of a large parity conserving term proportional to $\mathcal{Q}^2$, a small parity violating term (the interference term) proportional to $\mathcal{Q}\zeta$, and a negligibly small term on the order of $\zeta^2$. The interference term is proportional to a pseudoscalar quantity that depends only on the field geometry, the \emph{rotational invariant} $\kvec\cdot\Bfield$~\cite{Dounas-Frazer2011arXiv}. Thus the interference term vanishes if $\Bfield$ and $\kvec$ are orthogonal. One way to achieve a nonzero rotational invariant is to orient $\kvec$ along $\Bfield$ (fig.~\ref{fig:1}).

We calculate the transition rate when $\Bfield$ is sufficiently strong to resolve magnetic sublevels of the final state, but not those of the initial state. This regime is realistic since Zeeman splitting of the initial and final states are proportional to the nuclear and Bohr magnetons, respectively. In this case, the total rate is the sum of rates from all magnetic sublevels of the initial state: $R\rightarrow \sum_{M_i}R(M_i)$. When the fields are aligned as in fig.~(\ref{fig:1}), the transition rate is
\begin{equation}\label{eq:A^2}
R_{\pm} \propto \mathcal{Q}^2M_f^2/[I(I+1)]\pm 2\zeta\mathcal{Q}M_f/\sqrt{I(I+1)},
\end{equation}
where the positive (negative) sign is taken when $\kvec$ and $\Bfield$ are aligned  (anti-aligned), and we have omitted the term proportional to $\zeta^2$.

Reversals of applied fields are a powerful tool for discriminating APV from systematic effects. The interference term in~(\ref{eq:A^2}) changes sign when the relative alignment of $\kvec$ and $\Bfield$ is reversed, or when $M_f\rightarrow -M_f$. The \emph{asymmetry} is obtained by dividing the difference of rates upon a reversal by their sum:
\begin{equation}\label{eq:asymmetry}\textstyle
 (R_+-R_-)/(R_++R_-) = 2[\sqrt{I(I+1)}/M_f](\zeta/\mathcal{Q}).
\end{equation}
Reversals are sufficient to distinguish APV from many systematic uncertainties. Nevertheless,  there still exist systematic effects that give rise to \emph{spurious asymmetries}, which may mask APV. We consider two potential sources of spurious asymmetry: misalignment of applied fields, and stray electric and magnetic fields. A stray electric field $\Efield$ may induce $E1$-$E1$ transitions via the Stark effect~\cite{Bouchiat1974,Bouchiat1975}. When $\kvec$ and $\Bfield$ are misaligned ($\kvec\times\Bfield\neq \mathbf{0}$), Stark-induced transitions may interfere with the allowed transitions yielding a spurious asymmetry characterized by the following rotational invariant:  $(\kvec\times\Bfield)\cdot\Efield$. Because the Stark-induced transition amplitude may be nonzero when $F \neq I$, APV and Stark-induced asymmetries can be determined unambiguously by comparing transitions to different hyperfine levels of the final state.

We propose to measure the transition rate by observing fluorescence of the excited state, and assume that the transition is not saturated: $\mathcal{I}< \mathcal{I}_\text{sat}\equiv \Gamma/(4\pi\alpha\mathcal{Q})$. In this regime, fluorescence is proportional to the transition rate. The statistical sensitivity of this detection scheme is determined as follows: The number of excited atoms is $N_f=N_iR_{\pm}t \equiv N \pm N'$, where $N_i$ is the number of illuminated atoms, $t$ is the measurement time, and $N$ and $N'\ll N$ are the number of excited atoms due to parity-conserving and parity-violating processes. The signal-to-noise ratio is $\SNR=N'/\sqrt{N}$, or
\begin{equation}\label{eq:SNR}
\SNR = 8\pi\alpha\mathcal{I}\zeta\sqrt{N_it/\Gamma} =  2(\mathcal{I}/\mathcal{I}_\text{sat})(\zeta/\mathcal{Q})\sqrt{N_i\Gamma t}.
\end{equation}
The SNR is optimized by illuminating a large number of atoms with light that is intense, but does not saturate the $i\rightarrow f$ transition.

We now turn our attention to the two-photon 462 nm $5s^2\;{}^1S_0\rightarrow5s9p\;{}^1P_1$ transition in ${}^{87}$Sr ($Z=38$, $I=9/2$). The transition is enhanced by the intermediate $5s5p\;{}^1P_1$ state ($\Delta = 34$~cm$^{-1}$), and the parity-violating $E1$-$E1$ transition is induced by NSD APV mixing of the $5s9p\;{}^1P_1$ and $5s10s\;{}^1S_0$ states ($\omega_{af}=184$~cm$^{-1}$). We used expressions presented in ref.~\cite{Dzuba_arXiv} to calculate the NSD APV matrix element: $\Omega_{fa}\approx 10\kappa$~s$^{-1}$, where $\kappa$ is a dimensionless constant of order unity that characterizes the strength of NSD APV. The width of the transition is determined by the natural width $\Gamma=1.15\times10^7$~s$^{-1}$ of the $5s9p\;{}^1P_1$ state~\cite{Sansonetti2010}. Resolution of the magnetic sublevels of the final state requires a magnetic field larger than $2\Gamma/g\approx10$~G, where $g\approx 0.1$ is the Land\'e factor of the $F=I$ hyperfine level of the $5s9p~^1P_1$ state. The reduced matrix element $d_{ni}=5.4$~$ea_0$ can be determined from the width $\Gamma_n=2.01\times10^8$~s$^{-1}$ of the intermediate state~\cite{Sansonetti2010}, and we estimate that $d_{an}\approx ea_0$, $\mathcal{Q}_{fn}/d_{an}\approx\alpha/2$, and $\mu_{fn}\ll Q_{fn}$.  Then the APV asymmetry associated with this system is about $8\kappa\times10^{-9}$ when $M_f=1/2$. To estimate the SNR, we consider experimental parameters similar to those of ref.~\cite{Tsigutkin2009}: $N_i\approx 10^7$ atoms illuminated by a laser beam of characteristic radius $0.3$~mm.  Optimal statistical sensitivity is realized when $\mathcal{I} =\mathcal{I}_{\text{sat}}\approx6\times10^5$~W/cm$^2$. In this case, eq.~(\ref{eq:SNR}) yields $\SNR\approx \kappa \times 10^{-3}\sqrt{t/\text{s}}$. The saturation intensity corresponds to light power of about 2~kW at 462~nm. High light powers may be achieved in a running-wave power buildup cavity.  With this level of sensitivity,  about 300~hours of measurement time are required to achieve unit SNR. The projected asymmetry and SNR for the Sr system are similar to their observed counterparts in the most precise measurements of NSD APV in Tl~\cite{Vetter1995}.

%Another potential candidate for the DPS is the 557 nm $6s^2\;^1S_0 \rightarrow 6s8p\;^1P_1$ transition in ${}^{137}$Ba ($Z=56$, $I=3/2$). In this system, $E1$-$E1$ amplitudes may be induced by NSD APV mixing of opposite-parity states $6s8p\;^1P_1$ and $5d6d\;^3D_2$. This mixing relies on admixture of configuration $6p8p$ in the $5d6d\;^3D_2$ state, which is expected to occur at the level of a few percent~\cite{Kulaga2001,KozlovPC}.

Another potential candidate for the DPS is the 741 nm $7s^2\;^1S_0\rightarrow 7s7p\;^3P_1$ transition in unstable ${}^{225}$Ra ($Z=88$, $I=3/2$, $t_{1/2}=15$~days).  This system lacks an intermediate state whose energy is nearly half that of the final state; the closest state is $7s7p\;{}^1P_1$ ($\Delta \approx 14000$~cm$^{-1}$). Nevertheless, it is a good candidate for the DPS, partly due to the presence of nearly-degenerate opposite-parity levels $7s7p\;^3P_1$ and $7s6d\;{}^3D_2$ ($\omega_{af}=5$~cm$^{-1}$).  In this system, NSD APV mixing arises due to nonzero admixture of configuration $7p^2$ in the $7s6d\;{}^3D_2$ state~\cite{FlambaumPRAR1999}. Numerical calculations yield $\Omega_{fa}d_{an}/\omega_{af}\approx2\kappa\times10^{-9}$~$ea_0$~\cite{Dzuba2000}, $d_{an}=0.4$~$ea_0$, $d_{ni}=5.7$~$ea_0$, and $\Gamma=2.8\times10^6$~s$^{-1}$~\cite{Dzuba2006}. Like for the Sr system, we estimate that $Q_{fn}/d_{an}=\alpha/2$ and $\mu_{fn}\ll Q_{fn}$, yielding an approximate asymmetry of $3\kappa\times10^{-7}$. Laser cooling and trapping of ${}^{225}$Ra has been demonstrated~\cite{Guest2007}, producing about $N_i\approx 20$ trapped  atoms. When $\mathcal{I}=\mathcal{I}_\text{sat}\approx10^{8}$~W/cm$^2$,  the SNR is $\kappa\times10^{-2}\sqrt{t/\text{s}}$.  For a laser beam of 0.3 mm, the saturation intensity corresponds to light power of about 300~kW at 741~nm. These estimates suggest that unit SNR can be realized in under 3~hours of observation time.

In conclusion, we presented a method for measuring NSD APV without NSI background. The proposed scheme uses two-photon $J=0\rightarrow 1$ transitions driven by collinear photons of the same frequency, for which NSI APV effects are suppressed by BES. We described the criteria necessary for optimal SNR and APV asymmetry, and identified transitions in ${}^{87}$Sr and ${}^{225}$Ra that are promising candidates for application of the DPS.

%
% =====================================================================
% Acknowledgements
% =====================================================================
\acknowledgments The authors acknowledge helpful discussions with D.~P.~DeMille, V.~Dzuba, V.~Flambaum,
M.~Kozlov, and N.~A.~Leefer. This work has been supported by NSF.

\end{document}